\documentclass[11pt]{revtex4-2}
\usepackage{relsize}
\usepackage{braket}
\usepackage{hyperref}
\usepackage{amsmath,amsfonts,amssymb}
\hypersetup{dvips,dvipdfm,colorlinks=true,urlcolor=magenta,filecolor=magenta,linktoc=page,citecolor=red,linkcolor=blue,bookmarks=true}
\usepackage{graphicx,epstopdf}
\usepackage{bm}
\usepackage{hyperref}
\usepackage{amsmath,amsfonts,amssymb}

\hypersetup{dvips,dvipdfm,colorlinks=true,urlcolor=magenta,filecolor=magenta,linktoc=page,citecolor=red,linkcolor=blue,bookmarks=true}
\usepackage{graphicx,epstopdf}
\usepackage{array}

\newcounter{defcounter}
\newcolumntype{L}[1]{>{\raggedright\let\newline\\\arraybackslash\hspace{0pt}}m{#1}}
\newcolumntype{C}[1]{>{\centering\let\newline\\\arraybackslash\hspace{0pt}}m{#1}}
\newcolumntype{R}[1]{>{\raggedleft\let\newline\\\arraybackslash\hspace{0pt}}m{#1}}

\begin{document}
	
	\title{Is the emergent scenario in the early Universe a consequence of the quantization of the real scalar field?  \\
	}
	\author{Subhayan Maity\footnote {maitysubhayan@gmail.com}}
	\affiliation{Department of Mathematics, Jadavpur University, Kolkata-700032, West Bengal, India.}

	\author{Sujayita Bakra\footnote {sulukhejuri@gmail.com}}
	\affiliation{Department of Physics, Diamond Harbour Women University, West Bengal- 743388, India.}

	\begin{abstract}
		The emergent scenario of cosmic evolution is a topic of great interest in recent cosmology, especially because it describes a non-singular origin of the Universe, unlike the Big Bang models. This cosmic evolution pattern has already been established through the non-equilibrium thermodynamic prescription. But those models are phenomenological and require physical interpretation from the quantum field theory perspective. This work is an effort to search for a quantum field theoretical reason to justify the emergent nature of the Universe and the nature of cosmic evolution at the early phase of cosmic expansion.    
	\end{abstract}

	\maketitle
	
	
	
	
	\section{Introduction}
	The emergent model of the Universe is becoming a very popular and interesting cosmic evolution model nowadays. Even in an ever expanding cosmic model, the Universe can be assumed to start from an  emergent epoch \cite{Banerjee:2007qi,Bhattacharya:2016env,Bose:2020xml,Chakraborty:2014ora,Ellis:2002we,Ellis:2003qz,Guendelman:2014bva} with a non-zero constant scale factor($a_E$). It is a non-singular evolutionary model, unlike the Big Bang model where there is a singularity in the infinite past. In some previous works\cite{Maity:2021mlx,Maity:2022gby}, S.Maity in collaboration with S.Chakraborty has already  successfully demonstrated the existence of the emergent Universe under diffusion mechanism. The cosmic  Diffusion process is a consequence of the non-equilibrium thermodynamic evolution of the Universe.  The dissipation of the energy-momentum tensor of cosmic fluid is compensated by considering an interaction with a real scalar field\cite{Benisty:2017eqh,Benisty:2017lmt,Benisty:2017rbw,Benisty:2018oyy}. For suitable phenomenological choice of the time dependent scalar field leads to  the emergent nature of the Universe. But there is no field theoretical model to justify the diffusion mechanism. 
	\par Again the non-equilibrium thermodynamics in cosmic perspective is an outcome of the particle creation-annihilation process which can be described by the quantization of the Hamiltonian of the cosmic fluid.  
	\par This work is such an attempt to justify the possibility of the emergent Universe from the quantum field theoretical point of view. Here the cosmic fluid is assumed to be the real Klein - Gordon field (real scalar field) under flat FLRW space-time.  Then canonical quantization of the corresponding Hamiltonian is done and the conditions for complete and smooth quantized Hamiltonian are obtained. Then  the possibility of an emergent era has been determined from these conditions.

	\section{Dynamics of real scalar field in FLRW space-time and its quantization}
		In the framework of Einstein's general relativity, the evolution of the cosmic fluid is governed by Friedmann equations in a flat FLRW Universe $ds^2= d t^2- a^2(t)(dx^2 +dy^2 +d z^2)$, $ a$ is the time dependent scale factor of the Universe.  Considering the cosmic fluid as a barotropic fluid with barotropic index $W$, the trivial form of Friedmann equations are
	\begin{equation}
		3 H^2= k \rho, ~ 2\dot{H}+3H^2 = -kP , \label{1}
	\end{equation}
	where $H=\frac{\dot{a}}{a}$, Hubble parameter of the Universe. $P$ and $\rho$ are the thermodynamic pressure and energy density of the cosmic fluid respectively. Effectively the conservation equation of the fluid will be as,
	\begin{equation}
		\dot{\rho}+3(1+W)H \rho =0  \label{2}.
	\end{equation} 
	In this work, the dynamics of the cosmic fluid is aimed to explore following the ideas of quantum field theory. Here we start from representing the cosmic fluid by the Lagrangian of a real scalar field (Klein-Gordon field). It is the most trivial field with no charge symmetry. It holds neither a rigid nor a gauge symmetry.  In flat space-time (Minkowski geometry), the form of the Lagrangian density is given by,
	\begin{equation}
		\mathcal{L}= \frac{1}{2} \partial _{\mu} \phi \partial ^{\mu}\phi-\frac{1}{2} m^2 \phi ^2, \label{3}
	\end{equation} with $m$, the mass of the particle. $\phi^* =\phi$ is a real scalar field of space-time. This Lagrangian density is Lorentz invariant. This Lagrangian density yields the equation of motion (standard Klein-Gordon equation) as
	\begin{equation}
		(\Box ^2 +m^2) \phi =0,  \label{4}
	\end{equation}
	where $\Box^2=\partial_{\mu}\partial^{\mu}$, the D' Alembertier operator. The standard KG equation follows Lorentz covariance. 
	\par 	Under the curved space-time geometry, it can be modified in the form
	\begin{equation}
		\mathcal{L}= \frac{1}{2} \nabla_{\mu}\phi\nabla^{\mu}\phi -\frac{1}{2} (m^2+\zeta R) \phi ^2 , \label{5}
	\end{equation}
	$\nabla_{\mu}$ represents the covariant derivative. $R$ is the Ricci scalar and $\zeta$ is the coupling parameter. For minimal coupling, we have chosen $\zeta =0$ . This modified Lagrangian density is also a Lorentz invariant quantity. The Euler- Lagrangian equation yields the K-G equation from this Lagrangian density of equation (\ref{5}),
	\begin{equation}
		\left(\nabla_{\mu}\nabla^{\mu} + m^2 \right)\phi(X). =0 \label{6}
	\end{equation}
	$X=x^{\mu}=(t,\vec{x})$ is the space-time four - vector.
	For flat FLRW metric, equation (\ref{6}) takes the form
	\begin{equation}
		\ddot{\phi}+3H \dot{\phi} - \nabla^{\prime  ^2} \phi+ m^2 \phi =0 .\label{7}
	\end{equation}
	
	$H=\frac{\dot{a}}{a}$, the Hubble parameter.  $\vec{\nabla}^{\prime}=\frac{1}{a}\vec{\nabla}$. Let's consider the formal solution of the equation (\ref{7}) in terms of the Fourier transform $\phi (X)=\int d^3k \tilde{\phi}(K)e^{-i\int \tilde{K} dX}$ . $\tilde{K}=\tilde{k}^{\mu}=(k^0,\vec{k})$, the effective four momenta in this case.
	This modified KG equation does not hold Lorentz covariance. So it exhibits spontaneous Lorentz symmetry breaking.
	The term $3H\dot{\phi}$ causes the dissipation of energy from  the real scalar field like a damped oscillator. 
	\par Here we adopt the modification of the Lagrangian of the cosmic fluid field only due to the altered metric as a manifestation of the gravity. Here we haven't  imposed any extra term containing the signature of coupling between the real scalar field and the gravity. This approach is justified because we aimed to focus on the dynamics of the cosmic fluid under curved space-time.  
	
	\par From the formal solution, one can estimate from equation (\ref{7}),
		\begin{equation*}
		{K^0} ^2 +3i H K^0 -k^{\prime 2} - m^2 =0.  \label{a} 
	\end{equation*} 
	
	 \par Here we assume the slow variation of $k^0$ i.e. $
	\frac{\dot{k}^0}{k^0}<<1$.  $\vec{x}\rightarrow \vec{x}^{\prime}=a \vec{x}, \vec{k}\rightarrow \vec{k}^{\prime}=\frac{\vec{k}}{a}$. $\vec{x}^{\prime}, \vec{k}^{\prime}$ are the comoving space-coordinate and momenta respectively.Eventually $\vec{k}.\vec{x}= \vec{k}^{\prime}.\vec{x}^{\prime}$. 
	
		The expression of $K^0$ can be found as,
	\begin{equation*}
		K^0 = -\frac{3}{2}i H \pm \omega(t),  \label{b}
	\end{equation*}
	 Here $\omega(t)= \sqrt{|\omega_0 ^2 - \frac{9}{4} H^2|},  ~\omega_0=\sqrt{k^{\prime 2}+m^2}$ .

	The solution of this K-G equation (\ref{7}) is found as,
	
	\begin{equation}
		\phi(X)=\frac{1}{(2 \pi)^{\frac{3}{2}}} \int\frac{1}{\sqrt{2 \omega_0}} d^3k^{\prime}\left[ \mathcal{A}(\vec{k}^{\prime},t) e^{-i\int KdX} +\mathcal{A}^* (\vec{k}^{\prime},t)e^{+i\int KdX}  \right], \label{8}
	\end{equation} 
	where $ \mathcal{A}(\vec{k^{\prime}},t)=e^{-\frac{3}{2}\int Hdt} \alpha (\vec{k^{\prime }})$ and  $ \mathcal{A}^*(\vec{k^{\prime}},t)=e^{-\frac{3}{2}\int Hdt} \alpha^* (\vec{k^{\prime }})$. $\int KdX= \int \omega (t) dt - \vec{k}.\vec{x}$. 
	
	\par Generally, the comoving momenta $k^{\prime}$ is dependent on time. Hence $\omega_0$ is also time dependent.
	
	\par  The solution of K-G equation is the  superposition of infinite numbers of damped harmonic oscillators with momentum values ranging $-\infty <k^{\prime}< \infty$.

	\par For quantization of the Hamiltonian, this solution of the field will be replaced by the field operator,
	\begin{equation}
		\hat{\phi}(X)=\frac{1}{(2 \pi)^{\frac{3}{2}}} \int\frac{1}{\sqrt{2 \omega_0}} d^3k^{\prime}\left[ \hat{A}(\vec{k}^{\prime},t) e^{-i\int KdX} +\hat{A^{\dagger}} (\vec{k}^{\prime},t)e^{+i\int KdX}  \right].     \label{9}
	\end{equation}
	The Hamiltonian in this case will be in the  trivial form (similar  as in  the Minkowski space-time),
	\begin{equation}
		\hat{h}=\frac{1}{2}\int d^3 x^{\prime} \left[{\dot{
				{
					\hat{\phi}}}}^2+|\vec{\nabla}^{\prime}\hat{\phi}|^2\ + m^2 \hat{\phi} ^2 \right]  \label{10}
	\end{equation}
	The form of the Hamiltonian as in the  equation (\ref{10}) looks very trivial in this form but the explicit form of $\dot{\hat{\phi}}$  in FLRW metric adds some non -trivial terms in the Hamiltonian. 
	\par The form of the normal ordered Hamiltonian in this scenario can be found as,
	\begin{tiny}
		
		\begin{equation}
			:\hat{h}:=\int  \omega_0 ~ d^3 k^{\prime}\{ \hat{A}^{\dagger}(\vec{k}^{\prime},t)\hat{A}(\vec{k}^{\prime},t)\} +\left( \frac{9 H^2}{2} + i3H \omega \right ) \int  \frac{d^3k^{\prime}}{2 \omega_0} \hat{A}(\vec{k}^{\prime},t) \hat{A}(-\vec{k}^{\prime},t)  \ e^{-2i\int \omega dt}+  \left( \frac{9 H^2}{2} - i3H \omega \right ) \int  \frac{d^3k^{\prime}}{2 \omega_0} \hat{A}^{\dagger}(\vec{k}^{\prime},t) \hat{A}^{\dagger}(-\vec{k}^{\prime},t)  \ e^{+2i\int \omega dt}       \label{11} 
		\end{equation}
		
	\end{tiny}
	
	i.e.
	$:\hat{h}:=\int  \omega_0 ~ d^3 k^{\prime}\{ \hat{A}^{\dagger}(\vec{k}^{\prime},t)\hat{A}(\vec{k}^{\prime},t)\} +\sqrt{9H^2 \omega^2 +\frac{81}{4}H^4} ~e^{+ i \theta} \int  \frac{d^3k^{\prime}}{2 \omega_0} \hat{A}(\vec{k}^{\prime},t) \hat{A}(-\vec{k}^{\prime},t)  \ e^{-2i\int \omega dt}+  \sqrt{9H^2 \omega^2 +\frac{81}{4}H^4}~e^{- i \theta} \int  \frac{d^3k^{\prime}}{2 \omega_0} \hat{A}^{\dagger}(\vec{k}^{\prime},t) \hat{A}^{\dagger}(-\vec{k}^{\prime},t)  \ e^{+2i\int \omega dt}  $, with $\theta= \tan^{-1}(\frac{2w}{3H})$.
	\par 
	The first term in the right hand side $ \hat{h}_0 =\int   \omega_0 d^3 k^{\prime}\{ \hat{A}^{\dagger}(\vec{k}^{\prime},t)\hat{A}(\vec{k}^{\prime},t)\}$ of the equation (\ref{11}) is perfectly quantized with $\hat{A}^{\dagger}(\vec{k}^{\prime},t)$ and $\hat{A} (\vec{k}^{\prime},t)$ , the creation and annihilation operator respectively of the boson particle with comoving momenta $\vec{k}^{\prime}$.
	
	\begin{equation}
		\hat{A}^{\dagger}(\vec{k}^{\prime},t) \ket{0}= \ket{\vec{k}^{\prime}},\hat{A}(\vec{k}^{\prime},t) \ket{0}=0,  \label{12}
	\end{equation} 
	
	where $\ket{0}$ is vacuum state or zero particle state.  The corresponding number operator can be defined as
	\begin{equation}
		\hat{N}(\vec{k}^{\prime},t)=\hat{A}^{\dagger}(\vec{k}^{\prime},t)\hat{A}(\vec{k}^{\prime},t)    \label{13}
	\end{equation} 
	
	The second term in the expression of the Hamiltonian in the equation(\ref{11}) can not be quantized.

	In this case, we have several alternative restrictions on the parameters to  handle the non-trivial term and get the quantized form of the
	Hamiltonian operator.

	\paragraph{Case : $1$} \par
	For quantization of the Hamiltonian, one condition may be $H=0$ but $a\neq 0$.

	  Such condition is found to be satisfied at the origin of the emergent Universe\cite{Banerjee:2007qi,Bhattacharya:2016env,Bose:2020xml,Chakraborty:2014ora,Ellis:2002we,Ellis:2003qz,Guendelman:2014bva,Maity:2022noc,Maity:2022lbq,Mukherjee:2005zt,Mukherjee:2006ds,Beesham:2009zw,Paul:2020bje,Zhang:2013ykz,Paul:2015eja,Debnath:2017xcu,Paul:2018ppy,Debnath:2020bno,Debnath:2021ncz,Paul:2022dsb,Paul:2021lvb,Paul:2010jb,Paul:2011nw,Ghose:2011fk,Labrana:2013oca,Paul:2019oxo}. 
	\begin{equation}
		H\simeq 0, a\simeq a_E ~\mbox{when}  ~ t \rightarrow -\infty ,\label{14}
	\end{equation} 
	with $a_E$ is the scale factor of the Universe at the emergent epoch.
	Under this condition,  one has
	\begin{equation}
		\hat{A}^{\dagger}(\vec{k}^{\prime},t) \rightarrow \left[a_E\right]^{- \frac{3}{2}} \alpha ^{\dagger}(\vec{k}), \hat{A}^(\vec{k}^{\prime},t) \rightarrow \left[a_E\right]^{- \frac{3}{2}} \alpha ^(\vec{k}) . \label{15}
	\end{equation}
	Also the non-trivial terms vanishes at this epoch.
	Hence one may find the Hamiltonian in the form
	\begin{equation}
		:\hat{h}: = (a_E)^{-3} ~:\hat{h}_{\mbox{Minkowski}}: ,   \label{16}  
	\end{equation} 
	where $:\hat{h}_{\mbox{Minkowski}}:$ is the normal ordered Hamiltonian  of the real K-G field in Minkowski space-time. 
	
	Here the number operator will also be static with time,
	\begin{equation}
		\hat{N}(\vec{k},t)= (a_E)^{- 3} \hat{N}(\vec{k})_{\mbox{Minkowski}}.    \label{17}
	\end{equation}
	
	\paragraph{Case : $2$}
	\par 
	The other condition for the perfect  free field quantization of the  Hamiltonian is   $\mbox{Re}(\omega)=\frac{3}{2}H, H\neq 0$. Hence we have $\omega_0=\frac{3}{\sqrt{2}}H, K^{\prime}=\sqrt{\frac{9}{2}H^2-m^2}$. Consequently here, the system will be quantized with one particular time dependent energy eigenvalue $\omega_0=\frac{3}{\sqrt{2}}H$ and one momentum value $K^{\prime}=\sqrt{\frac{9}{2}H^2-m^2}$.
	
	 Hence  the particles created at different epochs from the vacuum state have different energy and different momentum value. So we have to label the spectrum also by $H$ value. 
	\begin{equation}
		\hat{\vec{P}} \ket{H}= \vec{k}\ket{H}=\sqrt{\frac{9}{2}H^2-m^2  } ~~~\hat{k}\ket{H} ,  \label{18}
	\end{equation} with $\hat{k}$ is the unit vector along the momentum vector.
	\begin{equation}
		:\hat{h}: \ket{H} =\frac{3}{\sqrt{2}}H \ket{H}.  \label{21}
	\end{equation}
	Here the number operator $
	\hat{N}(\vec{k}^{\prime},t)$ will be evolving with time as,
	\begin{equation}
		\hat{N}(\vec{k}^{\prime},t)=	 a^{- 3} \hat{N}(\vec{k}^{\prime})_{\mbox{Minkowski}}.  \label{20} 
	\end{equation}
	\paragraph{Case : $3$}

	 If $\mbox{Re}(\omega) \neq \frac{3}{2} H $ and $H$ is non-zero, then one may develop the quantization method analogs with the interacting field theory.  
	Here the explicit form of $\dot{\phi}$ can be written as
	\begin{equation}
		\dot{\hat{\phi}}(X)=e^{-\frac{3}{2}\int H dt} \left[\dot{\hat{\phi}}_0 - \frac{3}{2} H \hat{\phi} _0\right]=e^{-\frac{3}{2}\int H dt}\dot{\hat{\phi}}_0 - \frac{3}{2} H \hat{\phi}. \label{21}
	\end{equation} 
	$\hat{\phi}=e^{\pm\frac{3}{2}\int H dt} \hat{\phi}_0$  .
	
	\par 
	Consequently the form of Hamiltonian can be found as,
	\begin{equation}
		\hat{h}=\int d^3 x \frac{1}{2} \left[ \dot{\hat{\phi}}_0 ^2 +  |\vec{\nabla}^{\prime}\hat{\phi}_0|^2 + m^2  \hat{\phi}_0^2            \right] + i\frac{3}{2}\omega H\int d^3x \left[\dot{\hat{\phi}}_0 \hat{\phi}_0 + \hat{\phi}_0 \dot{\hat{\phi}}_0   \right]  \label{22}
	\end{equation}
	Evidently, the Hamiltonian contains two parts, 
	$\hat{h} = \hat{\mathcal{H}}_0 + \hat{\mathcal{H}}^{\prime} $. The first part $\hat{\mathcal{H}}_0 =\int d^3 x \frac{1}{2} \left[ \dot{\hat{\phi}}_0 ^2 +  |\vec{\nabla}^{\prime}\hat{\phi}_0|^2 + m^2  \hat{\phi}_0^2            \right]$ is the free field Hamiltonian and it can be quantized  with energy eigenvalue $\frac{\omega ^2}{\omega_0 a^3}$. But the second part $\hat{\mathcal{H}}^{\prime}$ can not be quantized and here we shall treat it as an interacting Hamiltonian. Hence we can calculate the $S$- matrix for this interacting Hamiltonian. 
	\begin{equation}
		\hat{S}= T e^{-i \int_{\mathcal{T}}^{t} \hat{\mathcal{H}}^{\prime}_I(t)dt},\label{23}
	\end{equation}  $T$ represents the time order product and $\hat{\mathcal{H}}^{\prime}_I(t)$  is the interacting Hamiltonian in the interaction picture representation. 
	\begin{equation}
		\hat{\mathcal{H}}^{\prime}_I(t)=  (i\frac{3}{2}\omega H)\int d^3x  \left[\dot{\hat{\phi}}_{0I} \hat{\phi}_{0I} + \hat{\phi}_{0I} \dot{\hat{\phi}}_{0I}   \right]. \label{24}
	\end{equation}
	Hence the spectrum at any epoch $t$ can be found from the relation 
	\begin{equation}
		\ket{\vec{k},H(t)}_I= \hat{S} \ket{\vec{k},H(\mathcal{T})}_I, \label{25}
	\end{equation}
	
	\par The system (the Universe) is truly free (the interacting part $\hat{\mathcal{H}}^{\prime}=0$) when $H=0 $ .
	 Again a valid free field Hamiltonian requires the conditions
	 $ H \simeq 0$ but $a\neq 0$.  Evidently, these conditions are found in the infinite past (the origin of the emergent Universe),
	 $t \rightarrow-\infty, H\simeq 0, a \rightarrow a_E$. The energy eigenvalue of the free Hamiltonian leads to $\frac{\omega_0}{a_E^3}$. In other epochs,  when  $\mbox{Re}(\omega) \neq \frac{3}{2}H$, the particle's state at any epoch can be related to the free field state (particle's state at $H\simeq 0$) through the $\hat{S}$ matrix. Let the free field state in the interaction picture  is denoted as 
	 \begin{equation}
	 	\ket{free}_I=\lim_{t\rightarrow{-\infty}}\ket{\vec{k},H}_I .  \label{26}
	 \end{equation}
	 
	 Hence the state at any arbitrary epoch $t$ with Hubble parameter $H$ can be found as,
	 \begin{equation}
	 	\ket{\vec{k},H}_I= \hat{S} \ket{free}_I, \label{27}
	 \end{equation}
	 
	 Here the form of $S$ matrix is as
	 \begin{equation}
	 	\hat{S}= T e^{-i \int_{-\infty}^{t} \hat{\mathcal{H}}^{\prime}_I(t)dt}.\label{28}
	 \end{equation}

	\section{ emergent scenario and cosmic evolution   }
	
	      The conditions for quantization of the cosmic fluid suggests that both the free field quantization and the interacting field quantization require the existence of an emergent scenario (when $H=0$). In both cases, the Hamiltonian gets static at $t \rightarrow -\infty $ with $a\rightarrow a _E, :\hat{h}:=\frac{1}{a_E ^3}:\hat{h}_{\mbox{Minkowski}}:$, and behave like a KG field in a Minkowski space-time.
	      \par At other epochs (when $H\neq 0$), the quantization is possible in the following ways.
	      \par $(1)$ ~~~
	       Free field quantization is possible only for one specific time dependent energy and momentum eigenvalue  $\epsilon=\frac{3}{\sqrt{2}}H$ , $k^{\prime}=\sqrt{\frac{9}{2}H^2 - m^2}$. 
	
	\par $(2)$ ~~~ The interacting field quantization is possible due to the interacting Hamiltonian ($\mathcal{H}^{\prime}$) proposed in the equation (\ref{24}) except $\epsilon = \frac{3}{\sqrt{2}}H$.
	
	\par One may expect the same cosmic evolution pattern from the dynamics of the cosmic fluid under  both quantization processes . In this work, we shall follow the free field quantization process of the fluid particle with energy $\epsilon=\frac{3}{\sqrt{2}}H$ to obtain the cosmic evolution pattern in this model.
	\par In an isolated Universe, the total energy will be conserved. Hence one may write the total energy
	\begin{equation}
   E= <\hat{N}> < \hat{h}> , \mbox{Constant}. \label{29}
	\end{equation}
	$< \hat{h}>=\bra{H}\hat{h}\ket{H}=\epsilon(H)$ is the average energy of a particle at any particular epoch when the value of the Hubble parameter is $H$. $\hat{N}=\bra{H}\hat{N}\ket{H}=N(H)=N_0 a^{-3}$. Here $N_0=<\hat{\alpha}^{\dagger}(H)\hat{\alpha}(H)>$, number operator in Minkowski space.

	\par When $H\neq 0 , \omega=\frac{3}{2}H$, we have $\epsilon(H)=\frac{3}{\sqrt{2}}H$. Hence One may find from the equation (\ref{29}),
	\begin{equation}
			H\dot{N}+\dot{H}N=0 .  \label{30}
	\end{equation} 
	
	Incorporating the expression of $N$ in the equation(\ref{30}), we get the evolution equation
		\begin{equation}
		\dot{H}-3H^2=0 . \label{31}
	\end{equation}
	
	 The solution of the equation (\ref{31}) yields
	 \begin{eqnarray}
	 	H=H_0\left(\frac{a}{a_0}\right)^3  \label{32}\\
	 	a=a_0\left[1-3H_0(t-t_0)\right]^{-\frac{1}{3}} ,  \label{33}
	 \end{eqnarray}
 where $a_0=a(t_0), H_0=H(t_0)$ and $t_0$ is a reference epoch of time. It is already mentioned that this solution will be valid at $H\neq 0$. Hence it will not reflect the continuous evolution from the  epoch of origin $t\rightarrow-\infty, a\simeq a_E, H=0$ to any $H\neq$ epoch.
 Clearly, there is a discontinuity immediately after the universe starts accelerating from  $H=0$.
 
     \par  In order to connect the evolution of $H=0$ phase and non-zero Hubble parameter phase, we have proposed a modified form of the scale factor and Hubble parameter as follows
     \begin{eqnarray}
     	a=a_E+\left(a_0-a_E\right) \left [ 1-3H_0(t-t_0)        \right]^{-\frac{1}{3}}  \label{34} \\
     H= H_0\left( 1-\frac{a_E}{a}   \right)\left[\frac{a-a_E}{a_0-a_E}\right]^3.    \label{35}	
     \end{eqnarray}
	
	The  proposed solutions (\ref{34}),(\ref{35}) satisfy the evolution equation (\ref{31}) in the limit $\frac{a_E}{a_0}<<1$. On the other hand, it satisfies the condition of emergent scenario as equation (\ref{14}).

	The choice of scale factor and Hubble parameter perfectly describes the cosmic evolution pattern under the quantization of cosmic fluid.

 Also at the limit $t\rightarrow t_0$, the scale factor starts to increase very rapidly and it satisfies the symptoms of the beginning of inflation or early time acceleration.

		\begin{figure}[h]
		
		\begin{minipage}{0.49\textwidth}
			\centering
			\includegraphics*[width=0.9\linewidth]{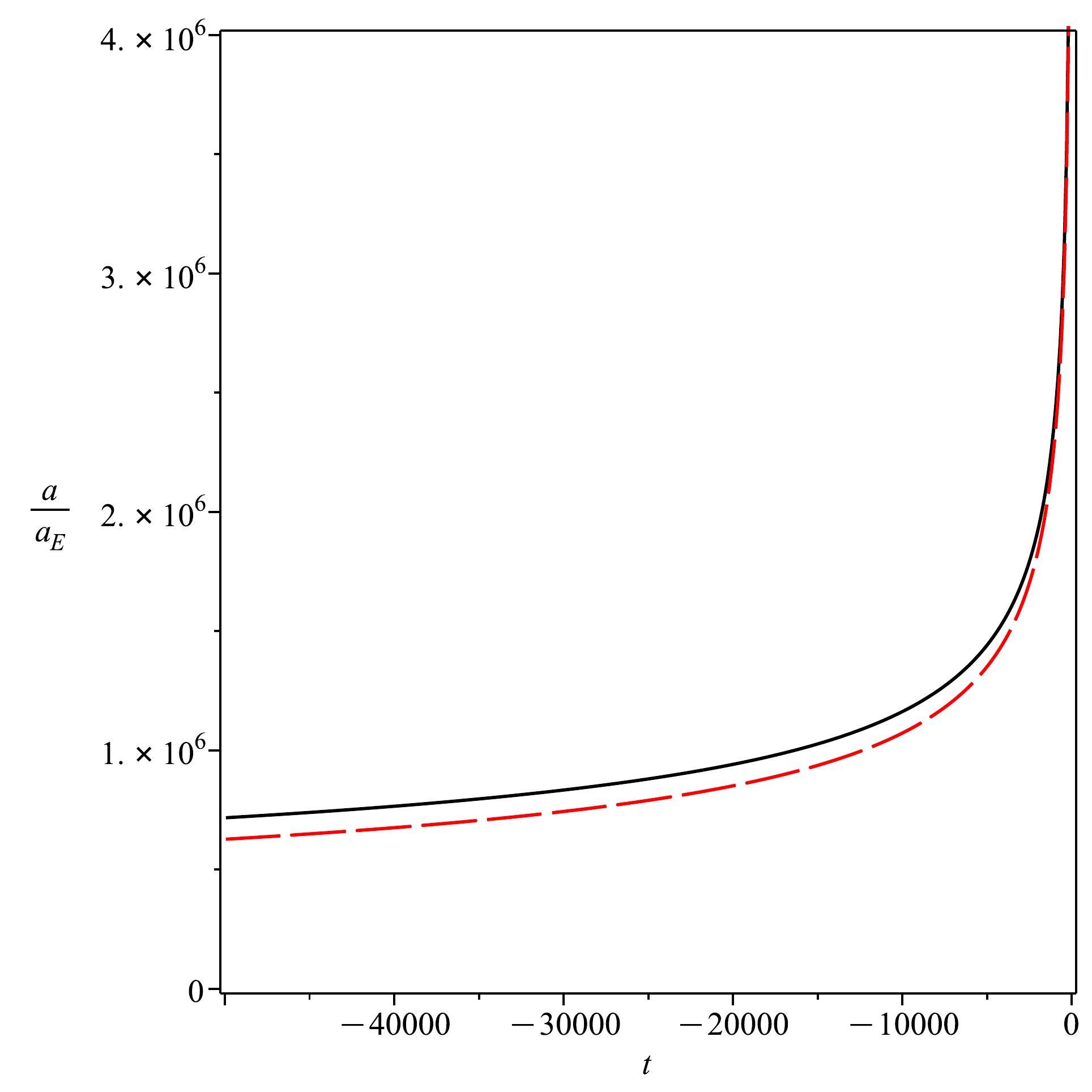}\\
			(a) 
		\end{minipage}
		\begin{minipage}{0.49\textwidth}
			\centering
			\includegraphics*[width=0.9\linewidth]{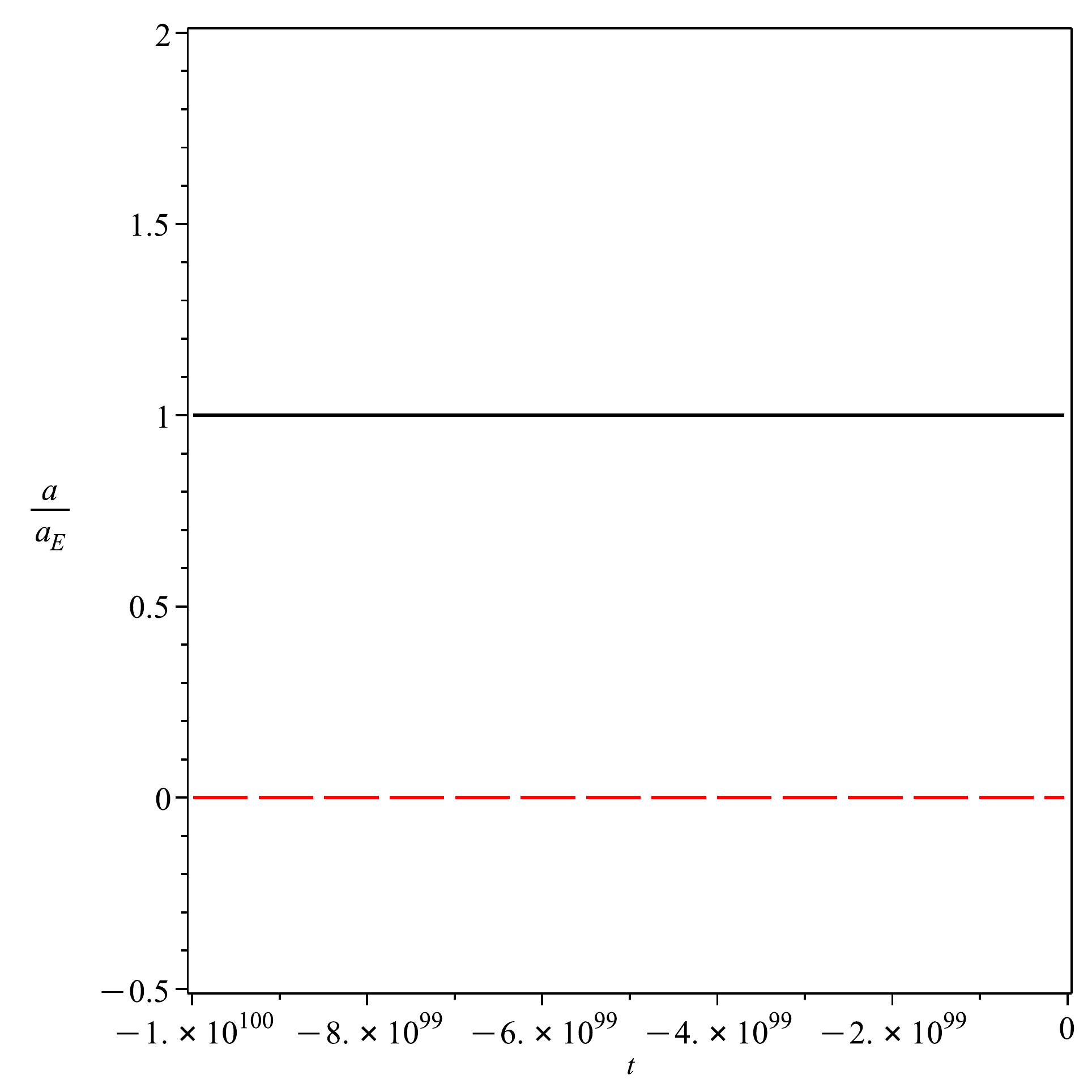}\\
			(b)
			
		\end{minipage}
	\begin{minipage}{0.49\textwidth}
		\centering
		\includegraphics*[width=0.9\linewidth]{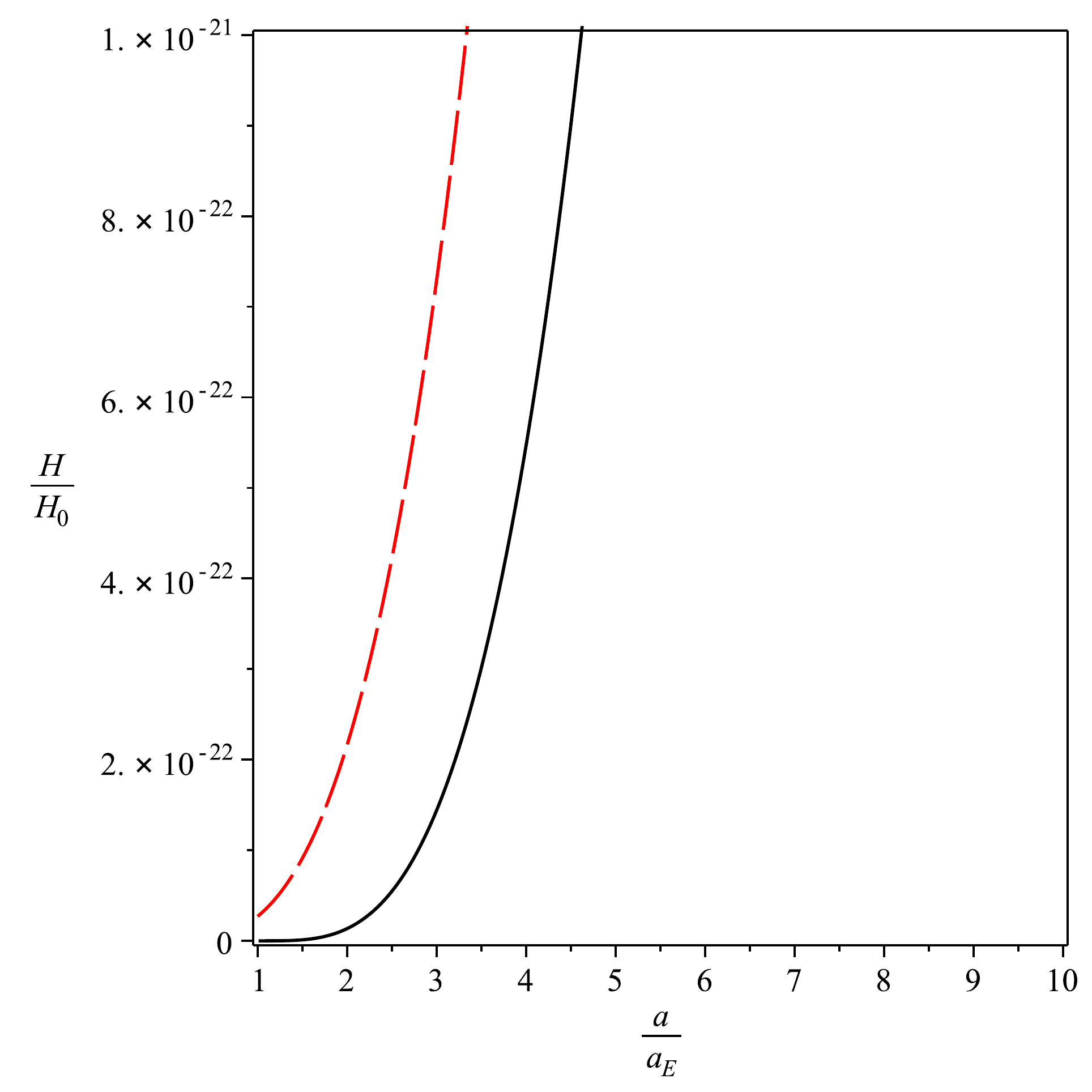}\\
		(c)
		
	\end{minipage}
		
		\begin{minipage}{0.76\textwidth}
			\caption{ Comparison between the solution (long dash, red) of equations (\ref{32}),(\ref{33}) and the proposed forms  (solid, Black) (\ref{34}) ,(\ref{35}) of the scale factor and the Hubble parameter .  $(a)$ In the top left, variation of scale factor with time  in the range of the time epoch $(-50000,0) $. $(b)$ In the top right, the variation of scale factor with time in the range of very early epoch of time $(-10^{100}, -10^{99})$. $(c)$ In the bottom, the variation of the Hubble parameter with the scale factor. Here we take $\frac{a_E}{a_0}=3 *10^{-8}, H_0=1$.}\label{fig1}
		\end{minipage}
	\end{figure}

	\begin{figure}[h]
		\begin{minipage}{0.49\textwidth}
			\centering
			\includegraphics*[width=0.9\linewidth]{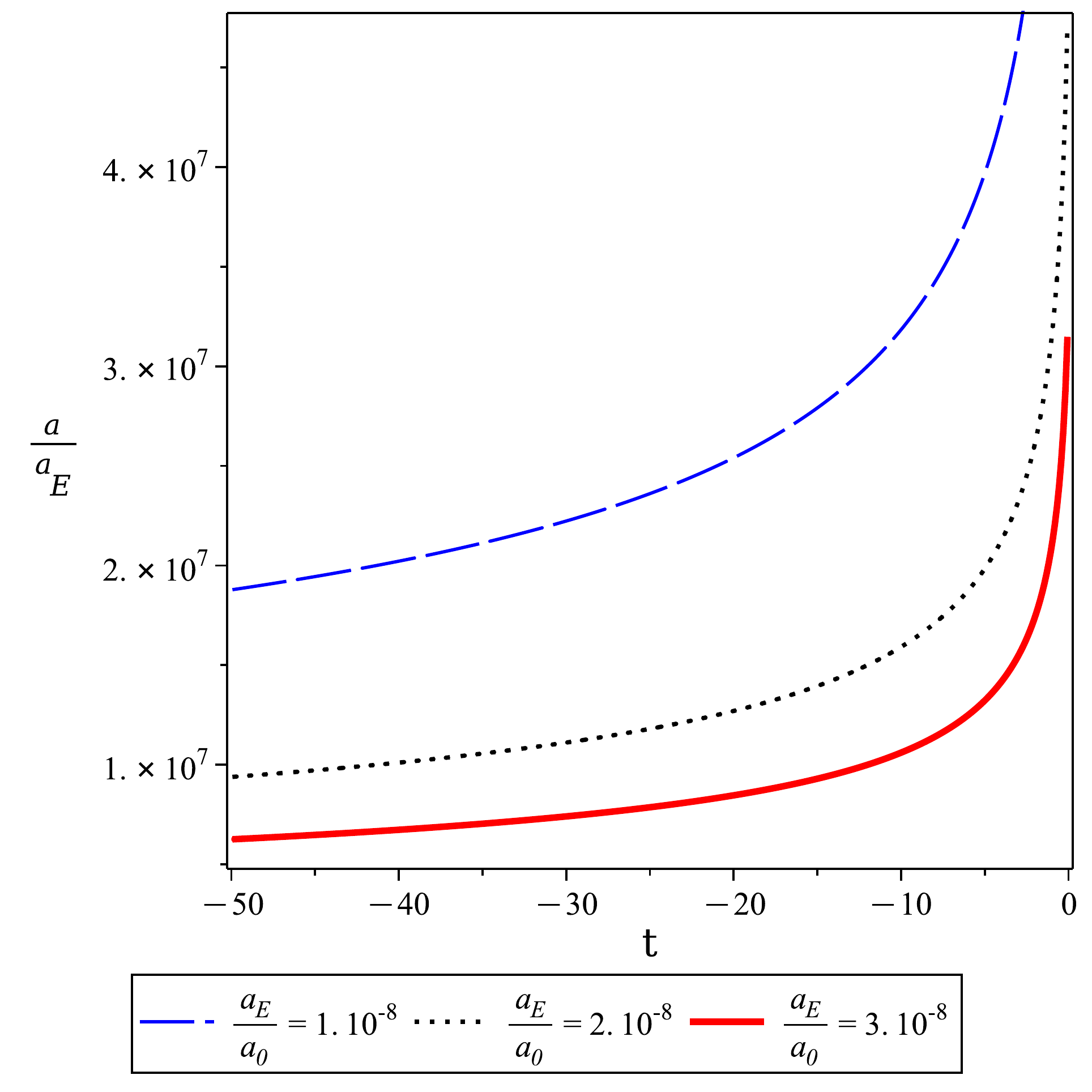}\\
			(a)
		\end{minipage}
		\begin{minipage}{0.49\textwidth}
			\centering
			\includegraphics*[width=0.9\linewidth]{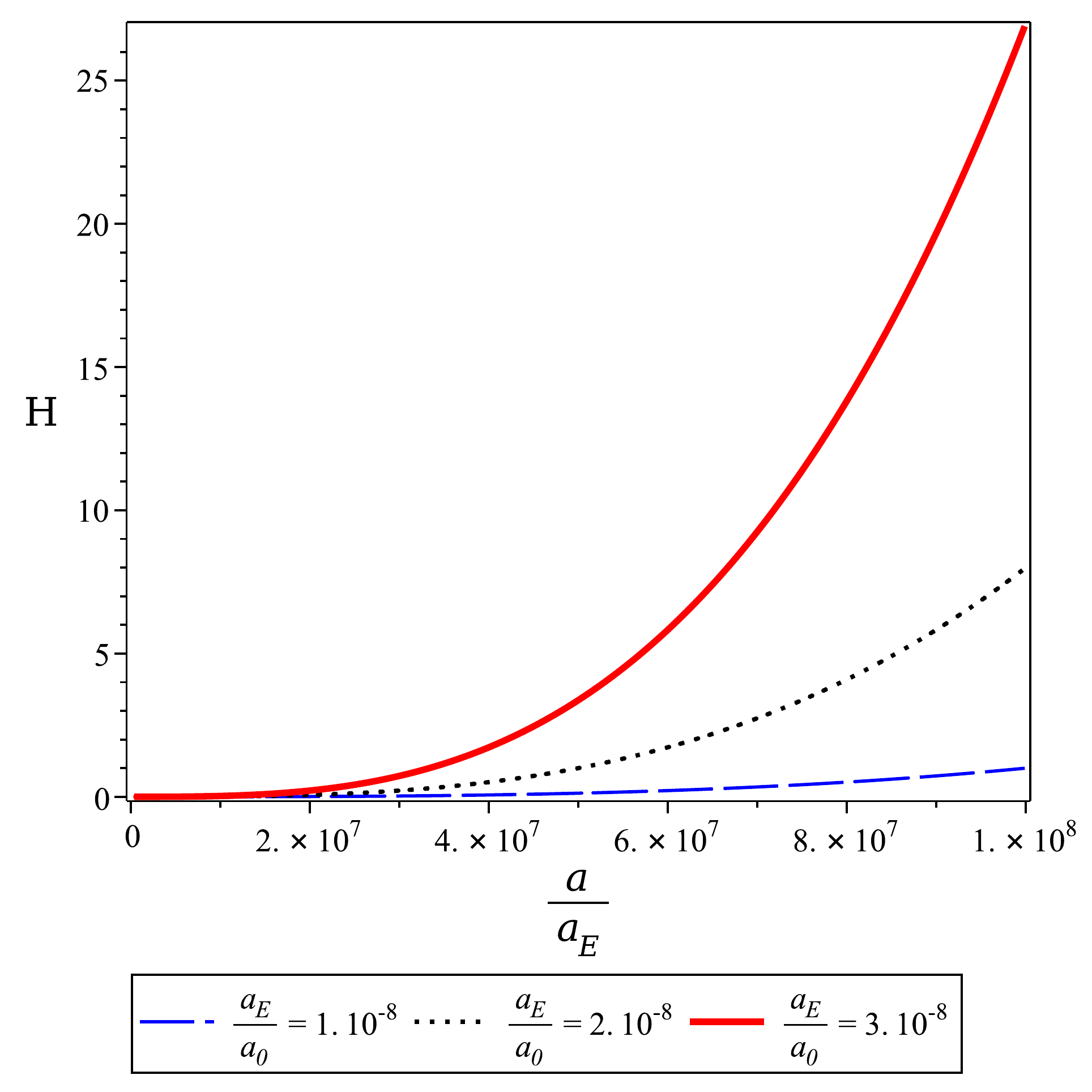}\\
			(b)
			
		\end{minipage}
		\begin{center}
			\caption{ (a)Variation of Scale factor  : $\frac{a}{a_E}$ with time $t$      (b) Variation of Hubble parameter : $\frac{H}{H_0}$with $\frac{a}{a_E}$,  for $t_0=0$ and three different values for the ratio $\frac{a_E}{a_0}$ for  $H_0=1$ .} 
		\end{center}
		\label{fig2}
		
	\end{figure}

	\begin{figure}[h]
		
		\begin{minipage}{0.49\textwidth}
			\centering
			\includegraphics*[width=0.9\linewidth]{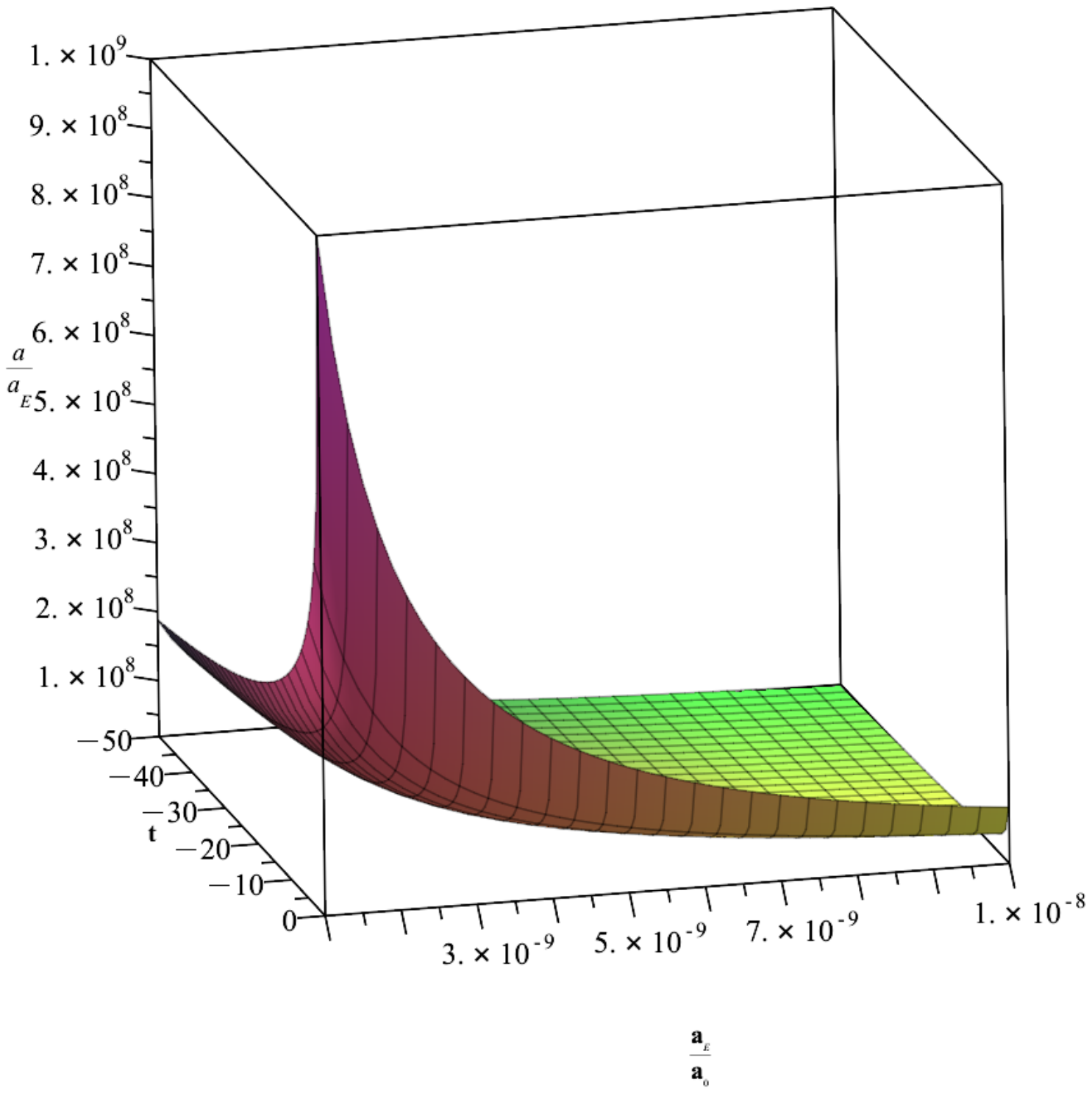}\\
			(a) 
		\end{minipage}
		\begin{minipage}{0.49\textwidth}
			\centering
			\includegraphics*[width=0.9\linewidth]{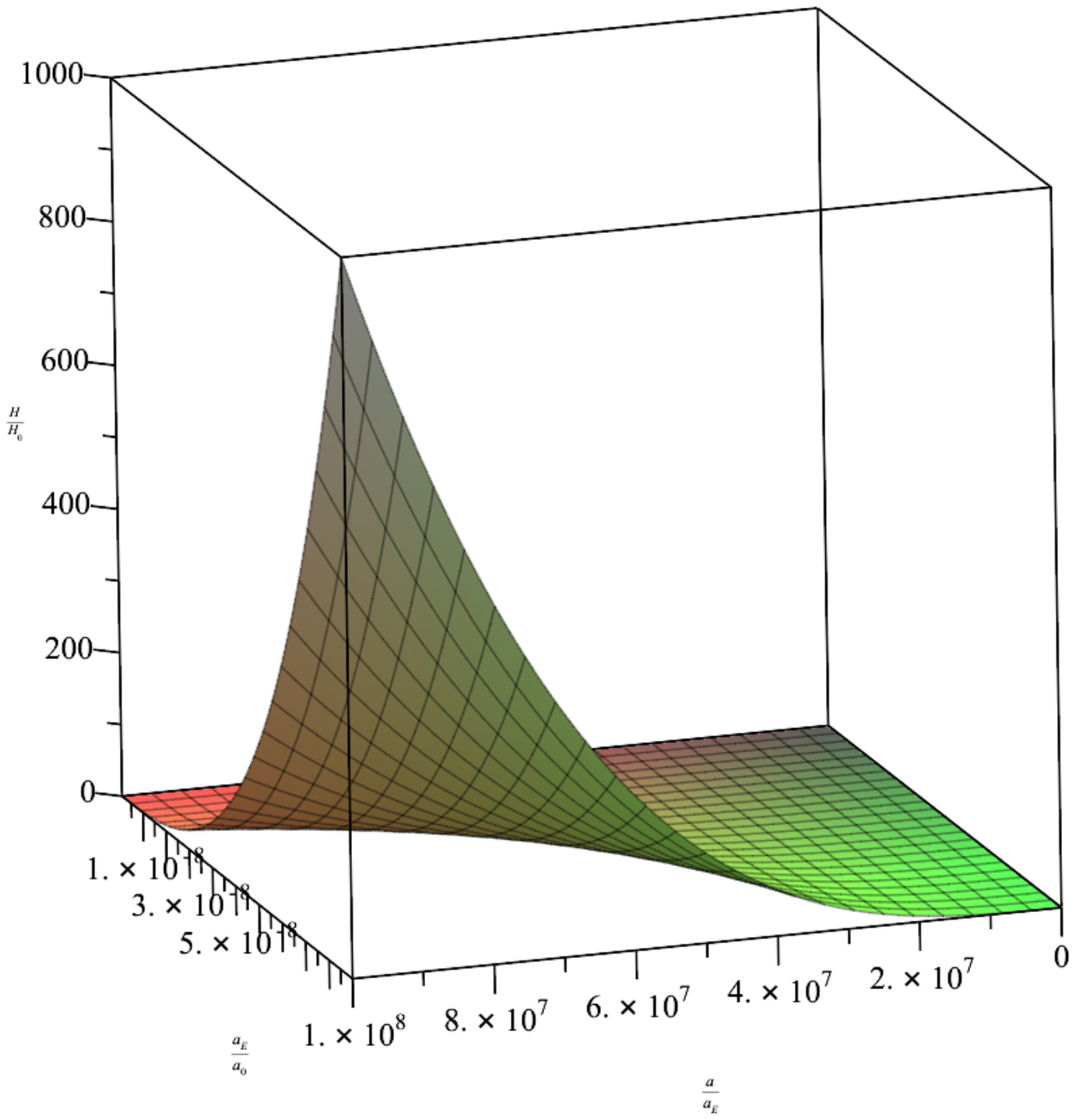}\\
			(b)
			
		\end{minipage}
		
		\begin{minipage}{0.76\textwidth}
			\caption{ (a) Evolution of  Scale factor :   $\frac{a}{a_E}$with time $t$ and values of $\frac{a_E}{a_0}$ (left) and (b) Variation of Hubble parameter :  $\frac{H}{H_0}$ with $\frac{a}{a_E}$ and $\frac{a_E}{a_0}$ (right) for $t_0=0$ in $3$d plots.}\label{fig3}
		\end{minipage}
	\end{figure}

	\par Reasonably one may consider  the reference epoch of time $t_0$ in the equation (\ref{34}) is the starting point of the inflationary era. Therefore the present model is the suitable description of the pre-inflationary cosmic evolution. From $t_0$, the cosmic evolution pattern will be changed and the evolution equation will have the different form. Thus the existence of mathematical Big-rip singularity of the solutions in equations (\ref{34}) and (\ref{35}) at $t=t_0 + \frac{1}{3H_0}$ has no physical relevance in this scenario.

	\section{Discussion } In this present model, we have successfully established the non-singular origin of the Universe from the perspective of quantum field theory.  Also we have shown two different alternatives for which the cosmic fluid field may be quantized at other epochs. As the conclusions, one gets that  the Universe consisted of the free real K-G field fluid at the emergent epoch. Then it starts evolving with particle creation process with a specific time dependent  energy  and momenta value or alternatively it may be evolved under an interacting Lagrangian mentioned (proposed) in the article(equation (\ref{24})). 
	\par Following the first option, the evolution pattern of the Universe has been obtained in the non-zero Hubble parameter epoch.

	    Also we have proposed a general evolutionary  scenario from the the epoch of the origin up to an inflationary phase. It is established that these proposed forms of the scale factor and the Hubble parameters fit well with the condition of quantization process and the cosmic evolution. 
	
	The deviation of the derived solutions (\ref{32}), (\ref{33}) from the modified solutions (\ref{34}),(\ref{35}) in Fig.$1$. It is found that these solutions are identical except the early epoch of time. They are going to coincide in the limit $a>>a_E$.
	
	 The proposed solutions(\ref{34}),(\ref{35}) are represented graphically in Fig.$2$ for three different values of the ratio $\frac{a_E}{a_0}$. In Fig.\ref*{fig3}
	, these parameters are shown in $3$d plots for better realization of the outcomes of this model. \par This work is a demonstration of finding the effect of traditional Friedmann cosmology from the dynamics of quantum fields.   
	
	\par Finally it is noteworthy that  it will be aimed to set up a convenient quantum field model to interpret the present late time acceleration along with a complete and continuous cosmic evolutionary scenario in a series of future works.
	\section*{Acknowledgment } The author SM acknowledges Prof. Subenoy Chakraborty, Dept. of Mathematics, Jadavpur University, Kolkata-$700032$ for his valuable suggestion on  this topic and the author SB thanks all the faculties, Dept. of Physics, Diamond Harbour Women University for their assistance while this work.

\end{document}